\documentstyle[twocolumn,aps,epsfig,graphics]{revtex}

\begin{document}

\title{A Novel Mechanism of \protect\protect\( H^{0}\protect \protect \) Di-baryon
Production in Proton-Proton Interactions\\
 from Parton Based Gribov-Regge Theory}

\author{M.~Bleicher\ddag, F.M.~Liu\ddag{}$^*$, J.~Aichelin\ddag, H.~J.~Drescher\dag, 
S.~Ostapchenko{}$^\%$, T.~Pierog\ddag\, and K.~Werner\ddag\vspace*{.5cm}}

\address{\ddag SUBATECH, Laboratoire de Physique Subatomique et des Technologies Associ\'{e}es
\\
 University of Nantes - IN2P3/CNRS - Ecole des Mines de Nantes \\
 4 rue Alfred Kastler, F-44072 Nantes Cedex 03, France}

\address{{}$^*$ Institute of Particle Physics , Huazhong
Normal University, Wuhan, China}

\address{\dag Physics Department, New York University,
New York, USA}

\address{{}$^\%$ Institut f\"ur Experimentelle Kernphysik,
University of Karlsruhe, Karlsruhe, Germany}

\address{{}$^\%$ Institute of Nuclear Physics,
Moscow State University, Moscow, Russia}

\maketitle
A novel mechanism of \( H^{0} \) and strangelet production in hadronic interactions
within the Gribov-Regge approach is presented. In contrast to traditional distillation
approaches, here the production of multiple (strange) quark bags does not require
large baryon densities or a QGP. The production cross section increases with center 
of mass energy. Rapidity and
transverse momentum distributions of the \( H^{0} \) are predicted for pp collisions
at \( E_{\textrm{lab}}=160 \)~AGeV and \( \sqrt{s}=200 \)~GeV. The predicted 
total \( H^{0} \) multiplicities are of order of  the \( \Omega ^{-} \) yield
and can be accessed by the NA49 and the STAR experiments. \vspace{.6cm}

The existence or non-existence of multi-quark bags, e.g. strangelets 
and (strange) di-baryons is one of the great open problems of 
intermediate and  high energy physics. 
Early theoretical models based on
SU(3) and SU(6) symmetries \cite{Oakes:1963,Dyson:1964} and on Regge theory
\cite{Libby:1969,Graffi:1969} suggest that di-baryons should exist. More recently,
QCD-inspired models predict di-baryons with strangeness S = 0, -1, and -2. The
invariant masses range between 2000 and 3000 MeV \cite{Jaffe:1976yi,Aerts:1978,Wong:1978,Aerts:1984ht,Kalashnikova:1987,Goldman:1989,Schaffner-Bielich:1999sy,Schaffner-Bielich:2000nd}.
Unfortunately, masses and widths of the expected 6-quark states differ considerably for these
models. However, most  QCD-inspired models predict di-baryons and none
seems to forbid them.  

Especially the search for a stable $H$-particle is closely related to the study of $\Xi$ 
and $\Lambda\Lambda$ hypernuclei (for very recent data on double $\Lambda$ hypernuclei 
see \cite{Ahn:sx,Takahashi:nm}). From observations on double $\Lambda$ hypernuclei, a mass
of $m_H> 2 m_{\Lambda\Lambda} - 28$~MeV is expected, while Jaffe estimated a binding 
energy of $\approx -80$~MeV.
The $H$ is a six quark state (uuddss) coupled to an SU(3) singlet in color and flavor.
Since its mass is smaller than $2 m_{\Lambda}$ it is stable against strong decays.
However, this object with baryon number two is not an ordinary nuclear state: the multi-quark
cluster contained in the $H$ is deconfined. Thus, the $H$ is the smallest strangelet
or might even be seen as a small droplet of Quark-Gluon-Plasma.
While on the hadronic side, hypernuclei are known 
to exist already for a long time, e.g. double \( \Lambda  \) hypernuclear events 
have been reported \cite{Ahn:sx,Takahashi:nm,Dalitz:1989}, no stringent 
observation of the $H$-particle exists. Even today, decades after the 
first prediction of the \( S=-2 \) $H$-di-baryon by 
Jaffe \cite{Jaffe:1976yi} the question of its existence is still open.

A major uncertainty for the detection of such speculative states is their
(meta)stability. Metastable exotic multihypernuclear objects (MEMOs), for example,
consists of nucleons, \( \Lambda  \)'s, and \( \Xi  \) and are stabilized
due to Pauli's principle, blocking the decay of the hyperons into nucleons. 
Only few investigations about the weak
decay of di-baryons exist so far (see \cite{Schaffner-Bielich:2000nd} for a full
discussion and new estimates for the weak nonleptonic decays of strange di-baryons):
In \cite{Don86}, the $H$-di-baryon was found to decay dominantly 
by \(H \rightarrow \Sigma ^{-}+p \)
for moderate binding energies. While the \( (\Lambda \Lambda ) \) bound state, which has
exactly the same quantum numbers as the $H$-di-baryon, was studied in \cite{Krivo82}.
Here, the main non-mesonic channel was found to be \( (\Lambda \Lambda ) \rightarrow \Lambda +n \).
If the life time of the \( (\Lambda \Lambda ) \) correlation or $H^0$ particle is not too long, 
the specific decay channels might be used to distinguish between both states. 

There are several searches in heavy-ion collisions for the $H$-di-baryon \cite{Belz96,Hank98}
and for long-lived strangelets \cite{Appel96,Arm97} with high sensitivities.
Hypernuclei have been detected most recently in heavy-ion reactions at the AGS
by the E864 collaboration \cite{Finch99}. 

In this letter we study the formation of the $H^0$-di-baryon within a new
approach called parton-based Gribov-Regge theory. It is realized in the 
Monte Carlo  program {\small NE}{\large X}{\small US} 3 \cite{Drescher:2000ha}. 
In this model  high energy hadronic
and nuclear collisions are treated within a self-consistent quantum
mechanical  multiple scattering formalism. Elementary interactions,
happening in  parallel, correspond to underlying microscopic (predominantly 
soft) parton  cascades and are described effectively as phenomenological
soft Pomeron exchanges. A Pomeron can be seen as a (soft) parton ladder,
which is attached to projectile and target nucleons via leg partons. 
At high energies one accounts also for the contribution of perturbative 
(high $p_t$) partons described by a so-called "semihard Pomeron" - a piece 
of the QCD parton ladder sandwiched between two soft Pomerons which are 
connected to the projectile and to the target in the usual way. The spectator 
partons of both projectile and target nucleons, left after Pomeron emissions, 
form nucleon remnants. 
The legs of the Pomerons form color singlets, such as q-\(\overline{\mathrm{q}} \),
q-qq or \( \overline{\mathrm{q}} \)-\( \overline{\mathrm{q}} \)\( \overline{\mathrm{q}} \).
The probability of q-qq and \( \overline{\mathrm{q}} \)-\( \overline{\mathrm{q}} \)\( \overline{\mathrm{q}} \)
is controlled by the parameter \( P_{\mathrm{qq}} \) and is uniquely 
fixed by the experimental yields on (multi-)strange baryons \cite{Liu:2002gw}. 

Particles are then produced from cutting the Pomerons and the decay of the remnants. 
As an intuitive way to understand particle production, each cut Pomeron 
is  regarded as two strings, i.e. two layers of a parton ladder.   
Each string has two ends which are quark(s) or antiquark(s) from the two 
Pomeron legs respectively.
To compensate the flavor, whenever a quark or an antiquark is taken as 
a string end, a corresponding anti-particle is put in the remnant nearby.

Since an arbitrary number of Pomerons may be involved, it is natural to
take quarks and antiquarks from the sea as the string ends.   
In order to describe the experimental yields on (multi-)strange baryons 
\cite{Liu:2002gw}, all the valence quarks stay in the remnants, 
whereas the string ends are represented by sea quarks.
Thus, Pomerons are vacuum excitations and produce particles and 
antiparticles equally\footnote{%
In addition to these singlet type processes, valence quark hard
interactions are treated differently in the present model.
To give a proper description of deep inelastic scattering data, a certain 
fraction of the Pomerons is
connected to the valance quarks of the hadron, not leading 
to a quark feeding of the remnant. 
This kind of hard processes will not be discussed here, but is included in 
the simulation.}. 
Only the remnants change the balance of particles and antiparticles,
due to the valence quarks inside.
Resulting in the possibility to solve the anti-omega puzzle 
\cite{Bleicher:2001nz} at the SPS.
\begin{figure}
 \par\resizebox*{!}{0.15\textheight}{\includegraphics{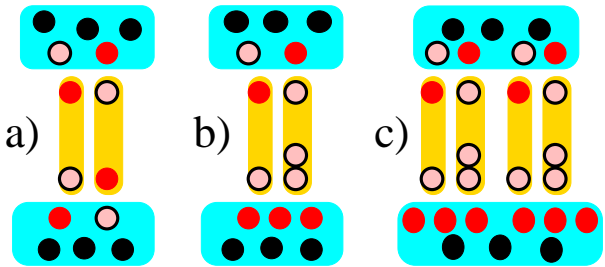}} \par{}
\vspace*{.3cm}
\caption{(a) The typical collision configuration has two remnants and one cut Pomeron
represented by two \protect\protect\( \mathrm{q}-\overline{\mathrm{q}}\protect \protect \)
strings. (b) One of the \protect\protect\( \mathrm{q}\protect \protect \)
string-ends is replaced by a \protect\protect\( \overline{\mathrm{q}}
\overline{\mathrm{q}}\protect \protect \) triplet state. 
To compensate the flavor, one of the remnants now has six quarks.
In the case of a uuddss flavor content a \protect\protect\( H^{0}\protect \protect \)
di-baryon can form. 
(c) Multiple Pomeron exchanges may lead to further accumulation of quarks in the remnant 
bag.
\label{quarkbag}}
\end{figure}

This prescription is able to accumulate quarks and di-quarks in the remnants
depending on the number of exchanged Pomerons. In the most simple case of a
single Pomeron exchange, the remnant may gain an additional di-quark and a
quark and is transformed into a six quark bag as discussed in the following.

The typical collision configuration has two remnants and one cut Pomeron represented
by two \protect\( \mathrm{q}-\overline{\mathrm{q}}\protect  \) strings, see
Fig. \ref{quarkbag}(a).

However, one or more of the \protect\( \mathrm{q}\protect  \) string-ends can be replaced
by a \protect\( \overline{\mathrm{q}}\overline{\mathrm{q}}\protect  \) triplet
state. To compensate the flavor, one of the remnants now has six 
quarks, cf. Fig.\ref{quarkbag}(b). This
possibility occurs with a probability \( P_{\mathrm{qq}} \). These six quarks are the
three valence quarks  u, u, d  plus three sea quarks, where each of
them may have the flavor  u , d, or  s, with relative weights
1 : 1 : \( f_{\mathrm{s}} \). Both parameters, \( P_{\mathrm{qq}} \) and \( f_{\mathrm{s}} \), are
uniquely determined by  multi-strange baryon data in proton-proton scatterings at
160~GeV to be \( P_{\mathrm{qq}}=0.02 \) and \( f_{\mathrm{s}}=0.3 \)\cite{Liu:2002gw}. Thus, there is a
small but nonzero probability to have a uuddss flavor in a remnant, such that a \( H^{0} \)
di-baryon may be formed. 
\begin{figure}
\vspace*{-.5cm}
\par \resizebox*{!}{0.3\textheight}{\includegraphics{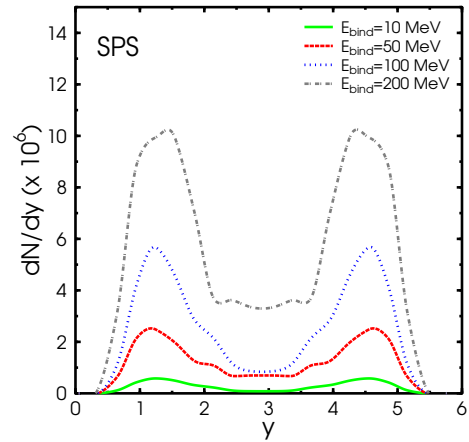}} \par{}
\caption{Rapidity distributions of \protect\protect\( H^{0}\protect \protect \)'s in
pp interactions at E\protect\( _{\textrm{lab}}=160\protect \)~GeV. The different
lines correspond to different binding energies: \protect\protect\( e_{\textrm{binding}}=10,50,100,200\protect \protect \)~MeV.
\label{dndy}}
\end{figure}
\begin{figure}
\vspace*{-.5cm}
 \par \resizebox*{!}{0.3\textheight}{\includegraphics{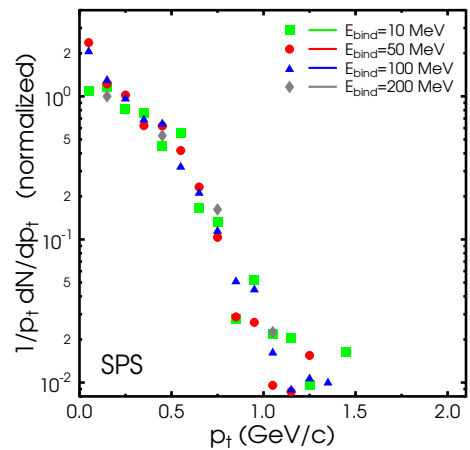}} \par{}
\caption{Transverse momentum distributions of \protect\protect\( H^{0}\protect \protect \)'s
in pp interactions at E\protect\( _{\textrm{lab}}=160\protect \)~GeV. The different symbols 
correspond to different binding energies: \protect\protect\( e_{\textrm{binding}}=10,50,100,200\protect \protect \)~MeV.
\label{dndpt}}
\end{figure}

The remnants have mass 
distribution \( P(m^{2})\propto (m^{2})^{-\alpha },\, \, \, m^{2}\in (m^{2}_{\mathrm{min}},\, x^{+}s), \)
here \( s \) is the squared CMS energy. With, \( m_{\mathrm{min}} \) being the minimal
hadron mass compatible with the remnant's quark content, and \( x^{+} \) is
the light-cone momentum fraction of the remnant which is determined in the collision
configuration. In the present study, the parameter \( \alpha  \) is 1.5 and
\( m_{\mathrm{min}}=2m_{\Lambda }-e_{\textrm{binding}} \). Remnants with masses
\( m \) between \( 2m_{\Lambda }-e_{\textrm{binding}}\: \mathrm{and}\: 2m_{\Lambda } \)
are considered to be \( H^{0} \) bound states. Projection on a specific 
spin state is omitted.

Contrary to the mechanism of \cite{Greiner:us}, which needs high baryon densities
to distill strangeness in heavy ion collisions, the present approach works differently:
It is independent of the baryon density and temperature. The presence of baryons enters
only due to multiple scatterings.
In addition, with increasing center-of-mass energy, multiple Pomeron exchanges gain importance.
This results in an increased possibility to produce heavy quark bags around the target and projectile
region of the collision as shown in Fig. \ref{quarkbag}(c)

Let us now study the multiplicities and momentum spectra of the calculated \( H^{0} \)'s.
Fig. \ref{dndy} depicts the rapidity distribution of the predicted \( H^{0} \)'s at the 
top SPS energy for a variety of possible binding energies. One observes a 
strong forward-backward peak in the \( H^{0} \) cross section indicating the 
production process from the remnants. 
\begin{figure}
\vspace*{-.5cm}
 \par \resizebox*{!}{0.3\textheight}{\includegraphics{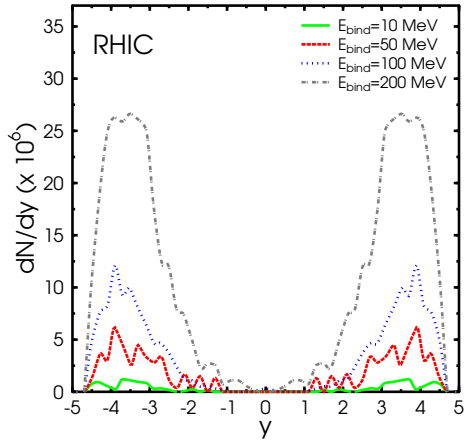}} \par{}
\caption{Rapidity distributions of \protect\protect\( H^{0}\protect \protect \)'s in
pp interactions at \protect\( \sqrt{s}=200\protect \)~GeV. The different lines
correspond to different binding energies: \protect\protect\( e_{\textrm{binding}}=10,50,100,200\protect \protect \)~MeV.
\label{dndy200}}
\end{figure}
%%%%%%%%%%%%%%%%%%%%%%%%%%%%%%%
\begin{table}
\begin{tabular}{crrr}
&$e_{\rm binding}$ (MeV) & Yield (SPS) $(\times 10^5)$ & Yield (RHIC) $(\times 10^5)$\\\hline
&10  & 0.141 & 0.313\\
&50  & 0.656 & 1.660\\
&100 & 1.351 & 3.424\\
&200 & 2.980 & 10.32\\
\end{tabular}
\caption{\label{table1} Predictions of the $H^0$ abundances in $4\pi$ 
for the different binding energies in pp collisions at 160~GeV (SPS) 
and $\sqrt s = 200$~GeV (RHIC).}
\end{table}
%%%%%%%%%%%%%%%%%%%%%%%%%%%%%%5

\noindent
Fig. \ref{dndpt} depicts the transverse momentum spectra of the \( H^{0} \)'s
for the same set of binding energies.

\begin{figure}
\vspace*{-0.5cm}
\par \resizebox*{!}{0.3\textheight}{\includegraphics{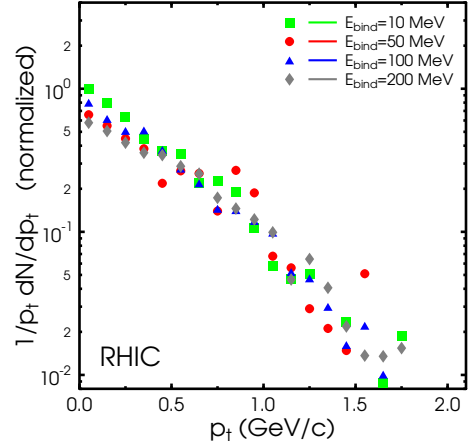}} \par{}
\caption{Transverse momentum distributions of \protect\protect\( H^{0}\protect \protect \)'s
in pp interactions at \protect\( \sqrt{s}=200\protect \)~GeV. The different
symbols correspond to different binding energies: \protect\protect\( e_{\textrm{binding}}=10,50,100,200\protect \protect \)~MeV.
\label{dndpt200}}
\end{figure}

In Figs. \ref{dndy200} and \ref{dndpt200} we show the rapidity and transverse
momentum spectra at \( \sqrt{s}=200 \)~GeV, again for the different binding
energies. Especially at RHIC energies one clearly observes the pile-up of
di-baryons in the forward and backward hemisphere. In the midrapidity region
the di-baryon yield vanishes.

\begin{figure}
\vspace*{-0.5cm}
\par \resizebox*{!}{0.3\textheight}{\includegraphics{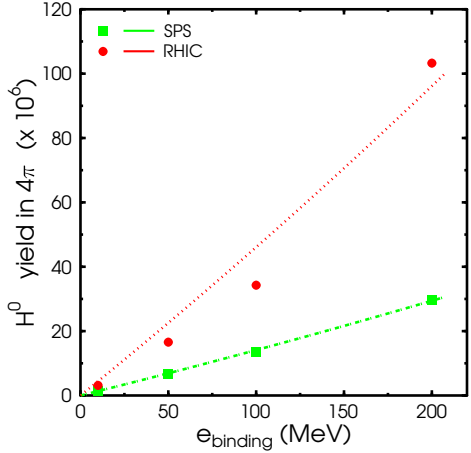}} \par{}
\caption{Integrated amounts of \protect\protect\( H^{0}\protect \protect \) as a function
of the binding energy for SPS (squares) and RHIC (circles) energies. 
The lines denote fits (see text). \label{integr}}
\end{figure}

We finally calculate the integrated abundances of \( H^{0} \) as a function of
the binding energy for SPS and RHIC energies, as shown in Fig. \ref{integr}.

\noindent
Since the total yields are given by 
\begin{eqnarray*}
\textrm{yield} & \propto  & \int ^{m_{\mathrm{max}}^{2}}_{m^{2}_{\mathrm{min}}}\, P(m^{2})\, {\rm d}m^{2}
% & = & \int ^{m_{\mathrm{max}}^{2}}_{m^{2}_{\mathrm{min}}}\, (m^{2})^{-\alpha }\, dm^{2}\\
=  \frac{m_{\mathrm{max}}^{2-2\alpha }-m^{2-2\alpha }_{\mathrm{min}}}{1-\alpha }\\
 & = & \frac{1}{m_{\Lambda }}\left[\frac{e_{\mathrm{binding}}}{2m_{\Lambda }}
+\left(\frac{e_{\mathrm{binding}}}{2m_{\Lambda }}\right)^{2}
+\left(\frac{e_{\mathrm{binding}}}{2m_{\Lambda }}\right)^{3}+\cdots \right]  \\
 & \cong & \frac{1}{m_{\Lambda }}\left[\frac{e_{\mathrm{binding}}}{2m_{\Lambda }}
+\left(\frac{e_{\mathrm{binding}}}{2m_{\Lambda }}\right)^{2}\right]
\end{eqnarray*}
where \( m_{\mathrm{max}}=2m_{\Lambda } \) , ~\( m_{\mathrm{min}}=2m_{\Lambda }-e_{\mathrm{binding}} \),
and \( \alpha =1.5 \). Thus, one finally expects a scaling of the multiplicities with
\begin{eqnarray*}
\textrm{yield}&\propto& e_{\mathrm{binding}}+\frac{e^2_{\mathrm{binding}}}{2m_\Lambda}\quad .  
\end{eqnarray*}

\noindent
Fig. \ref{integr} depicts a fit with \( \gamma \times (
e_{\mathrm{binding}}+ e^2_{\mathrm{binding}}/2m_\Lambda)
 \) and 
\( \gamma =1.35\times 10^{-7}\)~MeV$^{-1}$ (SPS, dashed line) and 
\( \gamma =4.41\times 10^{-7}\)~MeV$^{-1}$ (RHIC, dotted line), respectively.

The total abundances at both investigated collision energies and the set of binding energies 
are summarized in Table \ref{table1}: For optimistic
values of \( e_{\textrm{binding}} \), one expects 100 \( H^{0} \) to be observable
by NA49 in the next pp run \cite{addendum}. Compared to the predictions of
\cite{Schaffner-Bielich:1999sy} for nucleus-nucleus collisions at RHIC, the present 
study suggests additional di-baryon production
in the forward and backward hemisphere in addition to the midrapidity region.

In conclusion, we have presented a novel production channel of \( H^{0} \) di-baryons 
in pp collisions from parton-based Gribov-Regge theory. 
All model parameters are fixed by multi-strange baryon data at 160~GeV. 
In contrast to previous works, this mechanism does not require the
production of a deconfined state, neither does it need high baryon densities. 
In fact, one expects an increase in di-baryon production with energy strongly peaked in
beam direction. Note, that the suggested formation scenario of the $H^0$ (i.e. the 
lightest strangelet state), also invalidates strangelets as a smoking gun 
signature of a QGP state. Multiplicities, rapidity and transverse
momentum spectra are predicted for pp interaction at E\( _{\textrm{lab}}=160 \)~GeV
and \( \sqrt{s}=200 \)~GeV. 
At SPS, the cross section for \( H^{0} \) production in
the present study is found to be similar to the \( \Omega  \) production
cross section. Our predictions are  accessible in the $\Sigma^- p$ channel 
by the NA49 experiment at CERN and the STAR experiment at RHIC.

\vspace*{0.2cm}
\noindent
{\bf Error estimates}

\noindent
Table \ref{table2} shows the size of the analyzed samples for the different settings.
From this we obtain statistical errors on the multiplicities $3\% <\Delta N/N<9$\%.
%%%%%%%%%%%%%%%%%%%%%%%%%%%%%%%
\begin{table}
\begin{tabular}{rrrr}
&$e_{\rm binding}$ (MeV) & Events (SPS) & Events (RHIC) \\\hline
&10  & 100 Mio. & 50 Mio. \\
&50  &  50 Mio. & 20 Mio. \\
&100 &  50 Mio. & 50 Mio. \\
&200 &  10 Mio. & 12 Mio.\\
\end{tabular}
\caption{\label{table2} Analyzed events.}
\end{table}
%%%%%%%%%%%%%%%%%%%%%%%%%%%%%%5

\noindent
{\bf Acknowledgments}

\noindent
The authors want to thank J\"urgen Schaffner-Bielich for fruitful discussions on
the physics of strangelets and hyper matter. 
M.B. acknowledges support from the region Pays de la Loire.
S.O. acknowledges support by the German Ministry
for Research and Education (BMBF). H.J.D. kindly acknowledges 
support from NASA grant NAG-9246.

\end{document}